# Mutual-Excitation of Cryptocurrency Market Returns and Social Media Topics

Ross C. Phillips and Denise Gorse
Department of Computer Science,
University College London
Gower Street, London, WC1E 7JE, UK
e-mail: {r.phillips, d.gorse}@cs.ucl.ac.uk

*Abstract—* **Cryptocurrencies have recently experienced a new wave of price volatility and interest; activity within social media communities relating to cryptocurrencies has increased significantly. There is currently limited documented knowledge of factors which could indicate future price movements. This paper aims to decipher relationships between cryptocurrency price changes and topic discussion on social media to provide, among other things, an understanding of which topics are indicative of future price movements. To achieve this a well-known dynamic topic modelling approach is applied to social media communication to retrieve information about the temporal occurrence of various topics. A Hawkes model is then applied to find interactions between topics and cryptocurrency prices. The results show particular topics tend to precede certain types of price movements, for example the discussion of 'risk and investment vs trading' being indicative of price falls, the discussion of 'substantial price movements' being indicative of volatility, and the discussion of 'fundamental cryptocurrency value' by technical communities being indicative of price rises. The knowledge of topic relationships gained here could be built into a real-time system, providing trading or alerting signals.**

I. INTRODUCTION

Cryptocurrencies, of which Bitcoin is the most well-known, have recently experienced a new wave of interest. It has become commonplace to see TV coverage, news articles, blog posts, and discussion on social media platforms about cryptocurrencies. As well as mainstream excitement, there has been a flurry of activity from a range of interested parties: hedge funds have allocated funds for investment and trading within cryptocurrency markets; central banks are investigating the development of their own cryptocurrencies; consortiums have been set up to amalgamate the research efforts of otherwise competing companies into related (blockchain) technology. Although there has been increasing interest from a broad spectrum of groups, there is still little documented knowledge of how price movements can be predicted. Although positioned as currencies, research attempting to define what cryptocurrencies are—or at least why they are owned—notes that cryptocurrencies have many of the traits of speculative investments [1]. Given the current speculative nature of cryptocurrency market returns, one way to predict price changes might be to track changing interest in different cryptocurrency projects; social media, and other online indicators, offer one such way to track interest.

The link between social media and cryptocurrency markets has been demonstrated in the literature (discussed in Section II (A)) but also appears intuitive for a number of reasons. Primarily, cryptocurrency markets have historically, compared to traditional financial markets, been frequented more by home traders who—lacking both proximity to colleagues with whom they can discuss ideas and contractual obligations preventing them from doing so publically—may turn to social media. Furthermore, a lack of consistent regulation and anonymity of trader identities can cultivate an environment where people are likely to create social media activity around cryptocurrencies they own (informally termed in finance as "talking your book").

Given the widespread interest in trading cryptocurrency markets, knowledge of the discussion topics that affect price would be a useful component of any manual or automated trading strategy. The objective of this work is to explore the relationships between cryptocurrency market prices and social media discussion to understand what topics (and, indirectly, types of events) have the potential to predict price changes.

This work first retrieves occurrence of particular topics from social media content using dynamic topic modelling (an extension of Latent Dirichlet Allocation (LDA)) and then, using a Hawkes model, deciphers hidden interactions between topics and cryptocurrency market prices. A jump in topic discussion may cause further occurrences of the same topic (self-excitation) or occurrences of other topics or price changes (mutual-excitation).

This work also explores characteristics of different online communities (here, subreddits on the social media platform, Reddit), including how communication differs between communities and how activity within one subreddit may influence activity in other subreddits. More generally, this work is the first known application of a Hawkes model to both social media and financial data concurrently.

The remainder of this paper is structured as follows. Section II reviews relevant literature on the areas this work combines: cryptocurrency prediction via online data sources, topic modelling, and Hawkes models. Section III outlines the methodology used in this work. Section IV details the data retrieval process for both the social media and cryptocurrency market data. Section V details the experiment design. Section VI presents and discusses the results, and Section VII concludes.

## II. BACKGROUND

### A. Cryptocurrency prediction via online data sources

The use of online information, including social media, to predict financial asset movements has generated widespread interest. There exists a broad range of methods spanning from economics, data mining, natural language processing and machine learning to predict a range of financial assets [2]. As a subset of financial assets, prediction of cryptocurrency markets via online indicators has become popular. For example, Google searches for Bitcoin-related terms have been shown to have a relationship with the Bitcoin price [3]. It has been speculated that another indicator, relevant Wikipedia views, may provide a digital footprint of new users learning about a cryptocurrency [1]; such views exhibit a bidirectional relationship with price [3].

Twitter is a common source of social media data in the pursuit of financial asset prediction, and the prediction of cryptocurrency markets are no exception. Previous work has considered the strength and polarisation of opinions displayed in Twitter submissions (tweets) [4]. It was found that an increase in the polarisation of sentiment (disagreement of sentiment) preceded a rise in the price of Bitcoin. In other work, a system was built which categorised tweets into "positive", "negative" and "uncertain" based on matching words with pre-defined wordlist categories [5].

Although Twitter is the platform most commonly chosen as a data source for social media mining, it has disadvantages for the cryptocurrency domain. Firstly, it has been observed that there are large amounts of cryptocurrency related spam messages on Twitter [6], potentially hindering data mining attempts. Secondly, users discussing Bitcoin on Twitter appear to have different behaviours to the majority of users; users discussing Bitcoin choose just to discuss Bitcoin (and other cryptocurrencies) and not discuss other topics, suggesting other platforms may be better suited to them [7].

Recently, the social news aggregation and communication platform Reddit has been shown by the current authors to be a valuable source of information relating to cryptocurrency markets. Activity on Reddit was used to detect the epidemic-like spread of investment ideas beneficial in the prediction of cryptocurrency price bubbles [8]. The application of wavelet coherence further validated these findings; identifying that correlations between factors derived from Reddit and cryptocurrency prices strengthen in bubble-like regimes [9].

### B. Topic modelling

Topic modelling techniques have recently been applied to Bitcoin-related discussion sourced from a forum dedicated to Bitcoin (bitcointalk.org) [10, 11]. In the case of [10], the application of dynamic topic modelling (explained later in Section III (A)) allowed the evolution of topics, and terms within those topics, to be tracked over time. Results showed how discussion relevant to certain related technologies has changed over time. For example in one of the discovered topics manually labelled by the authors as relating to '*Bitcoin mining*' (the technical process by which blockchain transactions are validated), the term *CPU* was common earlier in the dataset, whereas terms relating to superior technology such as *GPU* increased in popularity over time.

The same data source (*bitcointalk.org*) has also been used in other topic modelling work [11]. Granger causality was applied to discovered topics to investigate whether there were relationships present between the occurrence of particular topics and statistics relating to Bitcoin. It was found that the topic related to *China* had a significant Granger causality with the Bitcoin price.

### C. Hawkes models and their applications

Hawkes processes [12] model situations where the occurrence of an event increases the probability of subsequent events. Since their introduction, they have been applied to model a range of event-based situations, including in early work the occurrence of earthquakes [13]. In more recent work, the application of Hawkes models within the separate fields of finance and social media has become popular.

Within finance, Hawkes models have been used to provide an understanding of a variety of dynamics—for example, the occurrence of financial contagion between different markets [14]. Recently, a Hawkes model has been applied to stock market returns and news article sentiment [15], in the first known application of a Hawkes model to the joint modelling of financial markets and news. Interactions between four types of events were considered: positive and negative market return events, and positive and negative news sentiment events. The methodology allowed for several findings, including positive (negative) returns being linked with positive (negative) sentiment.

Separately, Hawkes models have been used to model interactions between several social media sources. One relevant recent application considered the arrival of user submissions on three social media websites—Twitter, Reddit and 4chan—and achieved an understanding of the influence the different platforms have on one another in the propagation of political news [16].

Finally, the combination of a Hawkes model with topic modelling allowed examination of the self-excitation and mutual-excitation of regional discussion topics (originating from Los Angeles) on Twitter [17]. Topics were extracted from a corpus of submissions using non-negative matrix factorisation; when the proportion of a topic in a submission was above 0.1, this was classified as an occurrence of that topic. This classification allowed a time series of topic occurrences (for selected topics) to be generated, upon which a Hawkes model was applied to decipher hidden relationships between the topics. For example one relationship found was that topics relating to holidays preceded topics relating to basketball, but topics relating to basketball did not precede topics relating to holidays.

## III. METHODOLOGY

In this work, dynamic topic modelling is first applied to social media communication to decompose discussion into distinct topics. A Hawkes model is then applied to the resulting time series of topic occurrence alongside cryptocurrency price series to decipher relationships. The discussion below provides an overview of the methodologies used.

### A. Topic modelling

Topic modelling involves using statistical models to discover themes occurring within a corpus automatically; the aim is to find a distribution of words in each topic and the distribution of topics in each document. A *topic* can be considered as a probability distribution over a collection of words, e.g. a topic relating to *football* (soccer) is more likely to contain the words *goal* and *offside* than a topic relating to *cricket*. Since its introduction in 2003 [18], LDA has become a popular unsupervised learning technique for topic modelling. LDA assumes each document contains multiple topics to different extents. The generative process by which LDA assumes each document originates is described below:

1. Choose N ~ Poisson(ξ).
2. Choose θ ~ Dir (α).
3. For each of the N words $W_n$:
   a. Choose a topic $Z_n$ ~ Multinomial(θ).
   b. Choose a word $W_n$ from $p(W_n | Z_n, \beta)$, a multinomial probability conditioned on the topic $Z_n$.

Essentially, for each document, the number of words, N, to generate is chosen (step 1). The process then randomly chooses a distribution over topics, θ (step 2). Then for each word to be generated in the document, the process randomly chooses a topic, $Z_n$, from the distribution of topics (step 3a), and from that topic chooses a word, $W_n$, using the distribution of words in the topic (step 3b).

The variables of interest are $\theta_{d,k}$ (the distribution of topic $k$ in document $d$) and $\beta_k$ (the distribution of words in topic $k$). These are latent (hidden) parameters that can be estimated (for a particular dataset) via inference; for brevity, the details of the inference process are omitted here but can be found in [18] for LDA and [19] for dynamic topic models (discussed below). Inference allows for retrieval of per-document topic distributions and per-topic word distributions.

In standard LDA, there is no understanding of both the ordering of words within a document and the ordering of documents within a corpus. The set of topics that make up a particular document does not affect the set of topics that make up the next document. In some contexts, this may not be appropriate. For example, email threads, global news, or in the case presented here, messages on social media, all examples where there are likely to be temporal trends in topics discussed. As an extension to LDA, a *dynamic topic model* was introduced in 2006 [19]. In a dynamic topic model, there is still no understanding of the order of words in a document, but the order of documents in the corpus is now accounted for, meaning a sequentially organised corpus can be examined for evolving topics. To achieve this, data is divided into time slices over which topics can evolve. It is assumed topics appearing in one time slice are influenced by topics appearing in the previous time slice; a more detailed definition is provided in [19].

### B. Hawkes models

A comprehensive explanation of Hawkes models can be found for example in [12, 20]; the below provides an overview which focusses on how they are applied here. Hawkes models can be used to decipher the interaction dynamics between a group of *K* processes where the *K* processes can be considered as an implicit latent network; although connections between processes cannot be directly observed, the connections can be inferred from the temporal patterns of *events* (emissions) occurring on each process, *k*. Events are specified depending on the context; an event is roughly a jump in time series values, for example, a jump in market returns or a jump in discussion of a topic—a definition of the event types relevant to this work will follow in Section V (B). The occurrence of an event on a particular process can cause an impulse response (hence increasing the likelihood of further events) on a) that process (self-excitation) and b) on other processes (mutual-excitation). Given events occurring on a number of processes, the application of a Hawkes model can quantify previously hidden connections between the processes, applied here with the aim of deciphering how topics are related to one another, and how price changes are related to topic occurrence. After being fit to the data, the Hawkes model will contain weights representing the directional strength of any interaction between processes; these weights can be considered as the expected number of events on process B resulting from an event on process A.

Fig. 1 shows a demonstration of three example processes; market returns and two topics, X and Y.

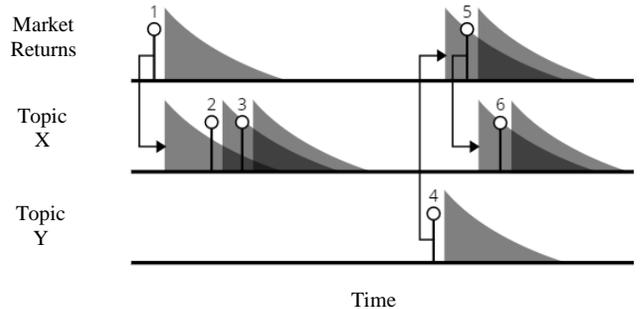

Figure 1. Example events (vertical line with open circle) and impulse responses (grey shading) on three processes (inspired by [20] and [16])

When relationships exist, they can be unidirectional or bidirectional. Each process has a background rate describing the rate of arrival of independent events (events not triggered by preceding events). Event 1 in Fig. 1 (a jump in market returns) is an example of an occurrence of an independent event. This causes an impulse response on its own time series (self-excitation) and on the Topic X time series (mutual-excitation). The impulse response increases the likelihood of events on these time series. Event 2 (a jump in topic discussion) occurs, and the resulting self-excitation prompts event 3. Event 4 then occurs on Topic Y's time series. This causes an impulse response on the market returns time series which prompts event 5; the resulting excitation prompts event 6. Overall, topic X appears responsive to price changes and Topic Y events appear to precede price events. The resulting weight $W_{Returns \to X}$ would be a number between 0 and 1 and $W_{X \to Returns}$ would be 0 implying no relationship in that direction. Other relationships can be deciphered similarly.

## IV. DATA SOURCES

### A. Social media

The social media platform Reddit attracts 8 billion page views per month. Reddit is separated into subreddits; there are over 50,000 active subreddits, where each subreddit is dedicated to discussion of a particular subject. Each major cryptocurrency project has an associated subreddit (and sometimes many). Table I displays the subreddits used in this work. For the cryptocurrencies considered, two subreddits for each are examined, one technical and one trading related. This allows for comparison between different community characteristics and their different interactions with the price.

TABLE I. SUBREDDITS CONSIDERED

| Cryptocurrency | Subreddit (Technical) | Subreddit (Trading) |
|---|---|---|
| Bitcoin | /r/Bitcoin | /r/BitcoinMarkets |
| Ethereum | /r/Ethereum | /r/EthTrader |

The previous work by the current authors [8, 9] has considered quantitative metrics to monitor user involvement within a particular subreddit such as the number of posts and new authors per day. To our knowledge, the current literature has not explored which topics are being discussed on cryptocurrency subreddits, which this work addresses by looking at the content of each submission. Every submission within a subreddit can be retrieved programmatically (in this case, with python scripts). The correct sequential ordering of submissions is maintained as each submission is timestamped.

### B. Cryptocurrency prices

The two largest cryptocurrencies (measured by market capitalisation) are used in this work: Bitcoin and Ethereum. All the required price data is sourced from the publicly available API of Bitfinex (a leading cryptocurrency trading exchange). Tick data (the most complete data possible) is retrieved, stored, and then aggregated to the required granularity. The experimental period chosen here is 30th August 2016 to 30th August 2017. During the data period used, the prices of both cryptocurrencies considered rose significantly, allowing us to study the interaction between prices and social media during this interesting period.

## V. EXPERIMENT DESIGN

### A. Topic modelling

As done commonly elsewhere (for example, in [18]), the corpus is pre-processed before topic modelling is applied. Stop words (commonly used words such as "the") are removed. Part-of-speech (POS) tagging is used to categorise words into types; nouns and adjectives are maintained while other types are removed. This filtering was decided upon based on preliminary work, and because it has been shown elsewhere that reducing a corpus to nouns can improve topic modelling results [21]. Finally, words appearing in less than 20 documents or more than 50% of documents are removed; such removals are commonly done elsewhere [19]. Once distinct topics have been identified by topic modelling, a time series of topic occurrence can be generated (if the proportion of a particular topic in a submission is above 0.1, this is classified as an occurrence of the topic, as in [17]).

A subset of topics are identified (and documented in Section VI (A)), based on their relevance and relative coherence (other less coherent topics are not analysed further). Highlighting only a subset of topics is common in topic modelling research (for example, in [10]). These chosen topics are then analysed in a Hawkes model, alongside market prices. This approach, to consider a subset of topics, has been used elsewhere when applying Hawkes models to topic modelling results [17]. This makes the assumption that the selected components (topics and market prices) exist in isolation (and ignores any explicit relationship with other factors not included in the model). This is suitable for the purpose of this analysis, to decipher how price changes relate to these chosen topics.

### B. Hawkes model

Hawkes models are most commonly applied to derived time series representing the occurrence of significant events (jumps/extreme changes) in the original time series, rather than to the original time series. The discussion below outlines the steps taken to identify such significant events.

Data is aggregated into fifteen-minute buckets ($\Delta t = 15$). This interval is small enough to avoid having too many overlapping events (for example, both a jump in market returns and a jump in topic occurrence) occurring within the same time bucket, and large enough to find mutual-excitation between buckets. Wider buckets (for example, one hour) are likely to group a number of events into a single bucket, losing the exact ordering of events. The same bucket size was chosen for similar work [15] after smaller intervals (2 and 5 minutes) failed to find excitation between processes. Instead of using absolute values (for example, the count of submissions containing a particular topic within that time bucket), log-returns between buckets are taken, as commonly done elsewhere (for example, [14]).

Jumps should be specified such that not every (non-zero) log-return is considered an event. A critical value is specified such that log-returns above this value are considered a jump, hence generating a time series of events to be considered by the Hawkes model. It was found that using $\Delta t = 15$ and the $99^{th}$ percentile of returns meant that 93% of events are non-overlapping (a similar percentage was seen in [16]). The maximum time for which an individual event can have an effect was chosen as one day (dt_max = 96 buckets). One reason for this choice is that cryptocurrencies are a globally traded market, and it takes time for news to propagate around the world. Experiments with variations of dt_max gave similar results.

Inference of parameters (based on the event-based data provided to the model) is achieved via Gibbs sampling, as detailed fully in [20].

## VI. RESULTS

### A. Topic modelling

Table II shows notable topics selected for their coherent cryptocurrency-related content. These topics have been manually labelled, as is common in topic modelling [11, 18]. The most probable words in each topic are retrieved from the final point in the dataset and displayed; although the probability of words (and thus the most probable words) within a topic varies gradually over time, the gist of the topic remains the same.

The balance between technical and non-technical discussion differs between Ethereum and Bitcoin. Of the 30 identified topics on each subreddit, /r/Ethereum contains only three topics that could be considered price or trading related, while /r/Bitcoin contains twelve. We hypothesise that this occurs for two reasons: 1) Bitcoin aims to be a currency (and hence price is a big part of it); 2) /r/Ethereum more actively discourages such discussion.

Many topics contain acronyms commonly used in cryptocurrency communities. For example, Topic 3 (on /r/Bitcoin) contains *btc* (Bitcoin), *eth* (Ethereum), *cap* (market capitalisation), *bch* (Bitcoin Cash), *btg* (Bitcoin Gold), and *ath* (all time high), while Topic 6 (on /r/Ethereum) contains *pow* (proof of work) and *pos* (proof of stake).

TABLE II. SELECTED TOPICS FROM EACH SUBREDDIT

| | # | Label | Most probable words |
|---|---|---|---|
| /r/Bitcoin | 1 | Mainstream adoption | site, dip, interested, website, Amazon, article, company, Google, group, page |
| | 3 | Trading terms / Bitcoin alternatives | btc, market, eth, cap, ratio, fork, trade, bch, btg, ath |
| | 20 | Substantial price movement | pump, moon[a], dump, sorry, list, quick, dude, random, it'll, way |
| /r/BitcoinMarkets | 4 | Downward price movement | big, crash, dip, bubble, huge, major, part, scam, correction, scale |
| | 17 | Risk / investment vs trading | trading, risk, everyone, worth, trade plan, way, advice, strategy, investment |
| | 26 | China / announcements | hope, statement, announcement, Chinese, list, announce, official, right, audit, illegal |
| /r/Ethereum | 6 | Consensus mechanisms | pos, pow, mining, stake, day, proof, security, network, mine, energy |
| | 8 | Hacks / Nervousness | money, attack, wait, someone, long, term, way, internet, short, iota |
| | 23 | Fundamental cryptocurrency value | value, crypto, currency, cash, eth, fiat, price, coin, market, news |
| /r/EthTrader | 4 | Trading terms | big, mean, moon[a], ratio, support, dip, break, chart, line, joke |
| | 7 | Future investments | money, next, crypto, real, devcon, link, investment, year, lot, half |
| | 24 | Mainstream adoption / app development | hope, private, key, site, Google, Amazon, bittrex, code, trust, app |

a. The term *Moon* may seem out of context, however it is cryptocurrency-specific jargon.

### B. Hawkes model

Fig. 2 (a) and Fig. 2 (b) show the strength of connections ($W_{vertical \rightarrow horizontal}$) between the considered processes for Bitcoin and Ethereum respectively. These weights are extracted from the Hawkes model after fitting to the dataset. Weights are displayed from the vertical to the horizontal axis; for example the bottom left cell on Fig. 2 (a) shows the weight value from Bitcoin price decrease (negative) events to /r/Bitcoin topic 1 events.

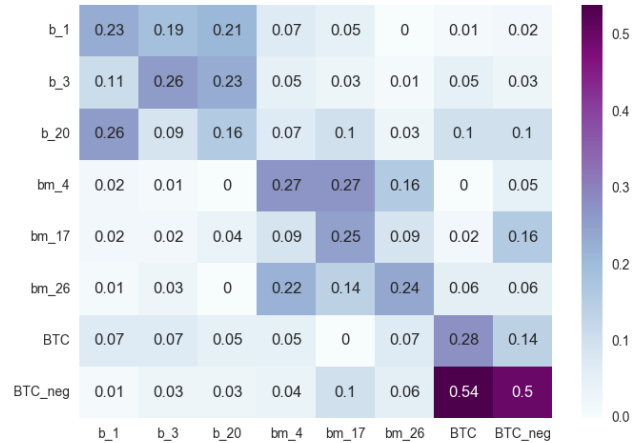

Figure 2(a). Weight values extracted from Hawkes model for Bitcoin-related topics and price movements; 'b_*' refers to the /r/Bitcoin subreddit and corresponding topic number '*' in Table II

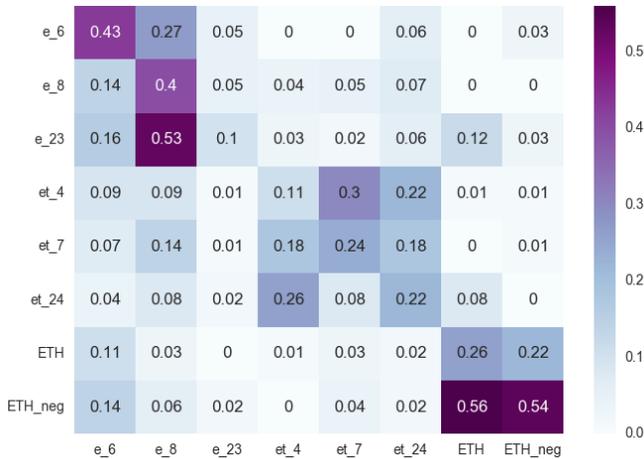

Figure 2(b). Weight values extracted from Hawkes model for Ethereum-related topics and price movements; 'e_*' refers to the /r/Ethereum subreddit and corresponding topic number '*' in Table II

There is a general pattern of stronger mutual-excitation between topics within a particular subreddit than between topics across different subreddits, observable by the cells having higher weights. Submissions are likely to prompt other submissions (in the same subreddit) as people reply to one another; these replies may contain a variety of different topics. Notable relationships are discussed below, starting with relationships between topics, then discussing the relationships observed involving market returns.

Occurrence of a particular topic can influence future occurrences of topics, to varying extents. Self-excitation (seen along the diagonal) is generally stronger than mutual-excitation (seen in non-diagonal cells), understandably as discussion of a topic is likely to prompt further discussion as people reply to each other. A significant non-diagonal relationship is the mutual-excitation between discussion of *'Downward price movement'* (/r/BitcoinMarkets topic 4) and discussion of *'Risk / investment vs trading'* (/r/BitcoinMarkets topic 17), evidenced by $W_{bm\_4 \rightarrow bm\_17} = 0.27$. It appears that jumps in discussion of downward price movements prompt people to discuss how they invested for the long-term (rather than actively trading) and are hence less sensitive to downward price movements. Elsewhere, discussion events relating to *'Fundamental cryptocurrency value'* (/r/Ethereum topic 23) are less likely to be triggered by other topics on the subreddit (seen by the smaller weights in the e_23 column), possibly because it is a distinct topic separate from the technical discussion.

It is however of greater interest to explore the relationships between topics and price movements. These relationships can be examined in the last two columns of each matrix. The topic *'Substantial price movement'* (/r/Bitcoin topic 20) has a stronger mutual-excitation with price events than most other topics, but doesn't indicate whether forthcoming price movements are positive or negative (as both $W_{b\_20 \rightarrow BTC}$ and $W_{b\_20 \rightarrow BTC\_neg} = 0.1$). This topic can hence be considered as being indicative of price volatility. *'Downward price movement'* (/r/BitcoinMarkets topic 4) has mutual-excitation with future negative price movements (based on $W_{bm\_4 \rightarrow BTC\_neg} = 0.05$) and has no relationship with future upwards price movement events. *'Risk / investment vs trading'* (/r/BitcoinMarkets topic 17) events has significant mutual-excitation with negative price movements; this, combined with the observation that negative price movements can excite this topic, suggests that, with a downward price movement users start to attempt to reassure others (and themselves) that they are invested for the long-term. However this topic appears to precede further price declines (based on $W_{bm\_17 \rightarrow BTC\_neg} = 0.16$, which is higher than any other topic-to-negative-price movement relationship for Bitcoin). Regarding the Ethereum-related subreddits, discussion of *'Fundamental cryptocurrency value'* (/r/Ethereum topic 23) has mutual-excitation with positive price increases (based on $W_{e\_23 \rightarrow ETH} = 0.12$ higher than any other topic to price movement relationship for Ethereum). While many topics on the trading subreddit relate to price movements and the value of Ethereum, very few topics on the technical subreddit relate to this, which might tend to add significance when it is discussed. Finally, discussion of *'Mainstream adoption/app development'* (/r/EthTrader topic 24) precedes price rise (versus price fall) events; this topic may relate to news or speculation of adoption, or may be indicative of overall positive sentiment.

As discussed briefly above (in relation to /r/BitcoinMarkets topic 17), price movements also influence topic discussion. Relationships between price and topics can be examined in the last two rows in each matrix. For example price increase events are likely to lead to discussion events of *'Mainstream adoption'* (/r/Bitcoin topic 1) (compared to price decrease events, which don't evidence such a strong relationship), as it is likely that if any news of such adoption comes out, the markets will react immediately and then the news will be discussed on social media for a period after it. In contrast both positive and negative price movements are likely to precede discussion of *'consensus mechanisms'* (/r/Ethereum topic 6). Ethereum developers are currently working on transitioning Ethereum from proof of work to proof of stake, so any news on this—positive or negative—can cause major price movements. In this case, the social media discussion events appear to lag price change events, suggesting the market is gaining awareness of this news from a source other than Reddit (possible alternative sources include GitHub progress and developers' Twitter accounts). Although topic discussion is lagging in this case, the topic is still strongly associated with price events. Intuitively, negative price movements are more likely than positive price movements to trigger discussion of *'Hacks / Nervousness'* (/r/Ethereum topic 8), in line with the observation $W_{ETH\_neg \rightarrow e\_8} = 0.06$ compared to $W_{ETH \rightarrow e\_8} = 0.03$). When a hack occurs it is likely the market will react quicker than social media, as demonstrated by the example to follow. On July 19th, 2017, an exploit was found in wallet software used by some to store their Ethereum, allowing an attacker to steal funds. Due to the uncertainty caused, the Ethereum price dropped approximately 15% over the first few hours. Social media discussion extended over 24 hours and beyond (first, news of the attack, then a few hours later actions taken to protect vulnerable wallets, then a post-mortem of the exploit published the next day—all causing events detected in this work).

Finally, price movements can influence the likelihood of future price movements. These relationships can be seen in the four cells at the bottom right of each matrix. For both Bitcoin and Ethereum there is strong self-excitation for both positive and negative returns; however for both cryptocurrencies, negative returns are more self-exciting than positive returns. This might result from: 1) negative returns inducing panic; 2) negative returns triggering stop losses, which can cause further negative returns. A Bitcoin price increase event is two times more likely to generate a further price increase event than generate a price decrease event (based on $W_{BTC \rightarrow BTC} = 0.28$ compared to $W_{BTC \rightarrow BTC\_neg} = 0.14$), a much larger ratio than seen for Ethereum, indicating Bitcoin is more trend following (for upwards price movements) than Ethereum.

## VII. CONCLUSIONS

This work considers a number of cryptocurrency-related discussion topics and the relationship between these topics and the price of the associated cryptocurrency. To achieve this, dynamic topic modelling was first applied to social media content, then a Hawkes model was used to decipher relationships between topics and cryptocurrency price movements. A number of topics were shown to precede price changes; for example the discovery that discussion of *'Fundamental cryptocurrency value'*, on the otherwise technical subreddit /r/Ethereum, precedes a positive return event; that discussion of *'Substantial price movement'* on /r/Bitcoin is indicative of price volatility; and that discussion of *'Risk / investment vs trading'* on /r/BitcoinMarkets, precedes a negative return event.

These discovered relationships could be built into a real-time trading or alerting system, the software infrastructure used here being reusable in such a system. Also, the methodology used could be expanded to a wider universe of cryptocurrencies. Data sources could be extended to other platforms, especially cryptocurrency-specific news websites, as it is likely the same topic-to-price relationships would exist elsewhere. A final natural extension to the research would be to consider the sentiment of subreddit discussion, while retaining a separation of the topic to which the sentiment relates. The aim would be to produce more accurate results than traditional sentiment techniques (which do not consider the topics involved).

REFERENCES

[1] F. Glaser, K. Zimmermann, M. Haferkorn, M.C. Weber and M. Siering, "Bitcoin - Asset or Currency? Revealing Users' Hidden Intentions" ECIS 2014
[2] M. Nardo, M. Petracco-Giudici and M. Naltsidis, "Walking down Wall Street with a tablet: a survey of stock market predictions using the web", Journal of Economic Surveys, vol. 30, no. 2, pp. 356-369, 2015.
[3] L. Kristoufek, "Bitcoin meets Google Trends and Wikipedia: Quantifying the relationship between phenomena of the Internet era", Scientific Reports, vol. 3, no. 1, 2013.
[4] D. Garcia and F. Schweitzer, "Social signals and algorithmic trading of Bitcoin", Royal Society Open Science, vol. 2, no. 9, 2015.
[5] J. Kaminski, "Nowcasting the Bitcoin Market with Twitter Signals", arXiv:1406.7577, 2014.
[6] M. Laskowski and H. Kim, "Rapid Prototyping of a Text Mining Application for Cryptocurrency Market Intelligence", 2016 IEEE 17th International Conference on Information Reuse and Integration (IRI), 2016.
[7] I. Hernandez, M. Bashir, G. Jeon and J. Bohr, "Are Bitcoin Users Less Sociable? An Analysis of Users' Language and Social Connections on Twitter", HCI International 2014 - Posters' Extended Abstracts, pp. 26-31, 2014.
[8] R. C. Phillips, D. Gorse. "Predicting Cryptocurrency Price Bubbles Using Social Media Data and Epidemic Modelling", IEEE Symposium Series on Computational Intelligence, 2017.
[9] R. C. Phillips, D. Gorse, "Cryptocurrency Price Drivers: Wavelet Coherence Analysis Revisited", PLOS ONE, 2018.
[10] M. Linton, E. G. S. Teo, C. Y. Chen and W. K. Härdle, "Dynamic topic modelling for cryptocurrency community forums", SFB Discussion Paper 2016
[11] Y. Kim, J. Lee, N. Park, J. Choo, J. Kim and C. Kim, "When Bitcoin encounters information in an online forum: Using text mining to analyse user opinions and predict value fluctuation", PLOS ONE, vol. 12, no. 5, 2017.
[12] A. Hawkes, "Spectra of Some Self-Exciting and Mutually Exciting Point Processes", Biometrika, vol. 58, no. 1, p. 83, 1971.
[13] L. Adamopoulos, "Cluster models for earthquakes: Regional comparisons", Journal of the International Association for Mathematical Geology, vol. 8, no. 4, pp. 463-475, 1976.
[14] Y. Aït-Sahalia, J. Cacho-Diaz and R. Laeven, "Modeling financial contagion using mutually exciting jump processes", Journal of Financial Economics, vol. 117, no. 3, pp. 585-606, 2015.
[15] S. Yang, A. Liu, J. Chen and A. Hawkes, "Applications of a multivariate Hawkes process to joint modeling of sentiment and market return events", Quantitative Finance, vol. 18, no. 2, pp. 295-310, 2017.
[16] S.Zannettou, T. Caulfield, E. Cristofaro, N. Kourtellis, I. Leontiadis, M. Sirivianos, G. Stringhini and J. Blackburn "The web centipede: understanding how web communities influence each other through the lens of mainstream and alternative news sources", 17th ACM Internet Measurement Conference. 2017.
[17] E. L. Lai, D. Moyer, B. Yuan, E. Fox, B. Hunter, A. L. Bertozzi and P. J. Brantingham. Topic time series analysis of microblogs. IMA Journal of Applied Mathematics, Volume 81, Issue 3, 2016
[18] D. Blei, A. Ng and M. Jordan, "Latent Dirichlet allocation", Journal of Machine Learning Research, vol. 3, pp. 993-1022, 2003.
[19] D. Blei and J. Lafferty. "Dynamic topic models". ICML, pp. 113-120, 2006.
[20] S. W. Linderman and R.P. Adams. "Discovering Latent Network Structure in Point Process Data", ICML, 2014.
[21] F. Martin and M. Johnson. "More Efficient Topic Modelling Through a Noun Only Approach", Australasian Language Technology Association Workshop, p. 111, 2015